\documentclass[11pt]{article}

\usepackage{amsfonts,amssymb}
\usepackage{amssymb}
\begin{document}

    \makeatletter \@addtoreset{equation}{section} \makeatother
    \renewcommand{\theequation}{\thesection.\arabic{equation}}
    \baselineskip 15pt

\newtheorem{defofentangidentical}{Definition}[section]
\newtheorem{defofentangidentical2}[defofentangidentical]{Definition}
\newtheorem{factorizabilityidentical2}{Theorem}[section]
\newtheorem{factorizabilityidentical3}[factorizabilityidentical2]{Theorem}
\newtheorem{anticomplex}{Theorem}[subsection]
\newtheorem{fermionschmidt}[anticomplex]{Theorem}
\newtheorem{measurefermion}[anticomplex]{Theorem}
\newtheorem{simcomplex}{Theorem}[subsection]
\newtheorem{bosonschmidt}[simcomplex]{Theorem}
\newtheorem{measureboson}[simcomplex]{Theorem}
\newtheorem{bequalb}[simcomplex]{Theorem}
\newtheorem{bdifferentb}[simcomplex]{Theorem}
\newtheorem{entropyin01}[simcomplex]{Theorem}

\title{\bf  Criteria  for  the  entanglement  of composite systems of
identical particles\footnote{Work  supported  in part  by
Istituto Nazionale di Fisica Nucleare, Sezione di Trieste, Italy}}

\author{GianCarlo Ghirardi\footnote{e-mail: ghirardi@ts.infn.it}\\
{\small Department of Theoretical Physics of the University of
Trieste, and}\\ {\small International Centre
for Theoretical Physics ``Abdus Salam'', and}\\ {\small Istituto
Nazionale di Fisica Nucleare, Sezione di Trieste, Trieste, Italy}\\ and \\
\\ Luca Marinatto\footnote{e-mail: marinatto@ts.infn.it}\\ {\small
Department of Theoretical Physics of the University of Trieste,
and}\\ {\small Istituto Nazionale di Fisica
Nucleare, Sezione di Trieste, Trieste, Italy}}

\date{}

\maketitle
\begin{abstract}
We identify a general criterion for detecting entanglement
 of pure bipartite quantum states describing a system of two identical
 particles. Such a criterion is based {\em both} on the consideration of 
 the Slater-Schmidt number of the fermionic and bosonic analog of the
 Schmidt decomposition {\em and} on the evaluation of the von Neumann entropy
 of the one-particle reduced statistical operators. 
\end{abstract}


\section{Introduction}

Entanglement is both one of the most peculiar and counterintuitive features 
 of Quantum Mechanics and one of the most valuable resources for Quantum
 Information and Quantum Computation.
In fact the possibility of implementing
 teleportation processes~\cite{tele}, of devising efficient quantum algorithms 
 outperforming the classical ones in solving certain computational
 problems~\cite{shor} and of exhibiting secure cryptographical
 protocols~\cite{ekert}, rests on the striking physical properties of 
 entangled states.
Accordingly, having an exhaustive knowledge of this phenomenon is
 extremely relevant both from the theoretical and from the practical point 
 of view.
However, despite that almost all the physical implementations of
 the above-mentioned processes developed so far involve identical particles,
 the very notion of entanglement in systems composed of
 indistinguishable elementary constituents seems to be lacking both of a 
 satisfactory theoretical formalization and of a clear physical understanding.
In fact, while when dealing with bipartite composite systems composed of 
 distinguishable particles and described by pure quantum states there exist 
 several equivalent criteria for revealing the entanglement
 (non-factorizability of the states, determination of the Schmidt number,
 evaluation of the von Neumann entropy of the reduced single-particle
 statistical operators), an uncritical application of these criteria to 
 quantum states describing a pair of identical particles seem to suggest 
 (mistakenly) that non-entangled states of identical particles cannot exist.  
This stems from the symmetrization postulate which, forcing the states of 
 identical particles to possess definite symmetry properties under any
 permutation of their labels, forbids the occurrence of 
 factorized states~\footnote{The only exception is represented by a system 
 of bosons described by the same state vector.}.
In order to clarify this subtle issue, where the unavoidable
 correlations arising from the truly indistinguishable nature of the
 particles are sometimes confused with the genuine correlations 
 due to the entanglement, we have been led to exhibit two (equivalent and,
 in our opinion, satisfactory) criteria for deciding whether a given 
 state is entangled or not.
The first is based on the possibility of attributing a {\em complete
 set of objective properties} to each component particle of the composed
 quantum system~\cite{gmw,gm}.
The second, whose analysis represents the main topic of this paper,
 is based {\em both} on the consideration of 
 the Slater-Schmidt number of the fermionic and bosonic analog of the
 Schmidt decomposition {\em and} on the evaluation of the von Neumann entropy
 of the one-particle reduced statistical operator~\cite{gm2}.
We will present briefly and schematically both criteria, 
 showing their complete equivalence and we address the interested reader
 to the original papers~\cite{gmw,gm,gm2} for a more exhaustive and 
 detailed treatment.


\section{Entanglement for distinguishable particles}

Let us first briefly review the basic features of 
 non-entangled (pure) state vectors associated 
 to composite systems of two distinguishable particles.
Given a bipartite state $\vert \psi(1,2) \rangle \in {\cal H}_{1} \otimes
 {\cal H}_{2}$, the following three equivalent statements represent
 necessary and sufficient conditions in order that the state
 can be considered as {\bf non-entangled}:
\begin{enumerate}
\item $\vert \psi(1,2) \rangle$ is factorized, i.e. there 
 exist two single-particle states $\vert \phi \rangle_{1} \in {\cal H}_{1}$ and
 $\vert \chi \rangle_{2} \in {\cal H}_{2}$ such that 
 $\vert \psi(1,2) \rangle = \vert \phi \rangle_{1} \otimes 
 \vert \chi \rangle_{2}$.
If this is the case a definite quantum state is assigned to each component 
 subsystem so that, since such states 
 are simultaneous eigenstates of a complete set of commuting observables,
 it is possible to predict with certainty the measurement outcomes of
 such a set of operators.
These outcomes represent objective properties which we can legitimately
 consider as possessed by the physical constituents.
\item The Schmidt number of $\vert \psi(1,2) \rangle$, that is the number
 of non-zero coefficients appearing in the Schmidt decomposition of the state,
 equals $1$.
\item If consideration is given to the reduced statistical operator 
 $\rho^{(i)}$
 of one of the two subsystems ($i=1,2$), its von Neumann entropy 
 $S(\rho^{(i)}) = - \textrm{Tr}\,[\, \rho^{(i)} \log \rho^{(i)}\,]$ equals 
 $0$~\footnote{For our convenience, the $\log$ function is intended
 to be in base $2$ rather than in the natural base $e$.}.
Since the von Neumann entropy measures the uncertainty about the quantum 
 state to attribute to a physical system, its value being null reflects the
 fact that, in this situation, there is no uncertainty at all concerning the 
 properties of each subsystem.
\end{enumerate}
The converse of the previous statements are easily proven.
Accordingly, a bipartite quantum system is described by an {\bf entangled}
 state $\vert \psi(1,2) \rangle$ if and only if one of the three following
 equivalent conditions holds true: i) the state is not factorizable; ii)
 the Schmidt number of the state is strictly greater than one; iii)
 the von Neumann entropy of both reduced statistical operators
 is strictly positive.

In this situation it is no more possible to attribute a definite quantum
 state to each constituent and therefore no objective property
 whatsoever can be claimed to be possessed by them; 
 accordingly, a positive value of the von Neumann entropy reflects this
 uncertainty concerning the state to attribute to the particles. 
 

\section{Two identical particles}

When passing to the case of composite quantum systems consisting 
 of identical particles, it has first of all to be noticed that 
 the symmetrization
 postulate poses severe constraints on the mathematical form of 
 their associated state vectors.
In fact, a couple of identical fermions or bosons must be described by
 antisymmetric and symmetric states respectively, states which are clearly not
 factorized in almost all cases.
Consequently their Schmidt decomposition involves more than one term
 and the von Neumann entropy of the reduced statistical operators is
 strictly positive.
It is therefore clear that, if one would resort to the same criteria used for 
 distinguishable particles in order to detect entanglement, one would be led to
 conclude that non-entangled states of identical particles cannot
 exist (with the only exception represented by two bosons in the same state).
To tackle the problem in the correct way, one has to stick to the idea
 that the physically most interesting and fundamental feature of non-entangled
 states is that {\em both constituents possess a complete set of objective
 properties}. 
In Refs.~\cite{gmw,gm} we have taken precisely this attitude, and we have given
 the following definitions:
\begin{defofentangidentical}
 \label{defofentangidentical}
 The identical constituents ${\cal S}_{1}$ and ${\cal S}_{2}$ of a
 composite quantum system ${\cal S}={\cal S}_{1}+{\cal S}_{2}$ are {\bf
 non-entangled} when both constituents possess a complete set of properties.
\end{defofentangidentical}
\begin{defofentangidentical2}
 \label{defofentangidentical2}
 Given a composite quantum system ${\cal S}={\cal S}_{1}+{\cal S}_{2}$ of
 two identical particles described by the normalized state vector
 $\vert \psi(1,2) \rangle$, we will say that one of the constituents
 possesses a complete set
 of properties iff there exists a one-dimensional projection operator $P$,
 defined on the single particle Hilbert space ${\cal H}$, such that:
\begin{equation}
 \label{wittgenstein}
 \langle \psi(1,2)\vert\,{\cal E}_{P}(1,2)\,\vert \psi(1,2)\rangle =1
\end{equation}
\noindent where
\begin{equation}
 \label{operatorediproiezione}
 {\cal E}_{P}(1,2)=P^{(1)}\otimes [\,I^{(2)}-P^{(2)}\,] +
 [\,I^{(1)}-P^{(1)}\,]\otimes  P^{(2)} + P^{(1)}\otimes P^{(2)}.
\end{equation}
\end{defofentangidentical2}
The second Definition is useful to make precise the meaning
 of the statement {\em ``both constituents possess a complete set of
 properties"} in the considered peculiar situation where it is not possible, 
 both conceptually and practically, to distinguish the two particles.
In fact condition of Eq.~(\ref{wittgenstein}) gives the probability of
 finding {\it at least} one of the two identical particles (without 
 saying which one) in the state associated to the
 one-dimensional projection operator $P$.
Since any state vector is a simultaneous eigenvector of a complete set of
 commuting observables, condition of Eq.~(\ref{wittgenstein}) allows to 
 attribute to {\em at least} one of the particles the complete set of 
 properties (eigenvalues) associated to the considered set of observables.

Starting from the above Definitions, we have proven the following
 Theorems~\cite{gmw,gm}:
\begin{factorizabilityidentical2}
 \label{factorizabilityidentical2}
 The identical fermions ${\cal S}_{1}$ and ${\cal S}_{2}$ of a composite
 quantum system ${\cal S}={\cal S}_{1}+{\cal S}_{2}$ described by the
 normalized state $\vert \psi(1,2) \rangle$ are {\bf non-entangled} iff
 $\vert \psi(1,2) \rangle$ is obtained by antisymmetrizing a factorized
 state.
\end{factorizabilityidentical2}
\begin{factorizabilityidentical3}
 \label{factorizabilityidentical3}
 The identical bosons of a composite quantum system ${\cal S}={\cal S}_{1}
 + {\cal S}_{2}$ described by the normalized state $\vert \psi(1,2)
 \rangle$ are {\bf non-entangled} iff either the state is obtained by
 symmetrizing a factorized product of two orthogonal states or it is
 the product of the same state for the two particles.
\end{factorizabilityidentical3}
These two Theorems clearly show why non-factorized state vectors describing
 two identical particles have not to be considered as necessarily entangled,
 since there exist cases in which property attribution is still possible
 in spite of their non-factorizable form.
In the situation described by Theorems~\ref{factorizabilityidentical2} 
 and~\ref{factorizabilityidentical3}, the states do not exhibit the 
 peculiar non-local 
 correlations between measurement outcomes which are typical of the 
 entangled states and, consequently, it is not possible to use them to violate 
 any Bell's inequality or to perform any teleportation process~\cite{gmw,gm}. 

 
\section{A new criterion to detect entanglement}

In the previous Section we have described a physically meaningful and
 unambiguous criterion for deciding whether a given state describing two
 identical particles is entangled or not.
The criterion is based on the possibility of attributing a definite 
 quantum state to each 
 component particle, without of course being able to specify which particle 
 has which property owing to their indistinguishability.
Recently other criteria for detecting entanglement have appeared in the 
 literature~\cite{cirac,pask,li,vacca}.
While some of them~\cite{cirac,pask} correctly deal with the case of identical 
 fermions, an inappropriate treatment of the (subtle) boson case is 
 presented in Ref.~\cite{pask} while in Ref.~\cite{li} the use of the entropy 
 criterion  is misleadig. 
Most of them~\cite{pask,li,vacca} seem to agree that the use of such criteria
 presents some obscure aspects.
The purpose of this Section is to exhibit a unified and unambiguous criterion, 
 based {\em both} on the consideration of 
 the Slater-Schmidt number of the fermionic and bosonic analog of the
 Schmidt decomposition {\em and} on the evaluation of the von Neumann entropy
 of the one-particle reduced statistical operators, to identify whether
 a state is entangled or not.
 
As we will show, such a criterion turns out to be in complete agreement
 with the previous criterion based on the property attribution and
 it clarifies all the obscure aspects which have been pointed out by
 the authors of Refs.~\cite{pask,li,vacca}. 

For the sake of simplicity we will deal separately with the cases of two
 identical fermions and two identical bosons.


\subsection{The fermion case}

The notion of entanglement for systems composed of two identical
 fermions has been discussed in Ref.~\cite{cirac} where a 
 {\em ``fermionic analog of the Schmidt decomposition''} has been exhibited.
Such a decomposition results from a nice extension to the set of the
 antisymmetric complex matrices of a well-known theorem holding
 for antisymmetric real matrices and it states that:
\begin{fermionschmidt}
 \label{fermionschmidt}
 Any state vector $\vert \psi(1,2)
 \rangle$ describing two identical fermions of spin $s$ and,
 consequently, belonging to the antisymmetric manifold
 ${\cal A}(\mathbb{C}^{2s+1}\otimes \mathbb{C}^{2s+1})$, can be written as:
\begin{equation}
 \label{fermion0.5}
 \vert \psi(1,2)\rangle = \sum_{i=1}^{(2s+1)/2} a_{i}\cdot
 \frac{1}{\sqrt{2}} \,[\,\vert 2i-1\rangle_{1} \otimes \vert 2i
 \rangle_{2}- \vert 2i \rangle_{1} \otimes \vert 2i-1 \rangle_{2}\,],
\end{equation}

\noindent where the states 
 $\left\{\, \vert 2i-1 \rangle, \vert 2i \rangle \right\}$
 with $i=1\dots (2s+1)/2$ constitute an orthonormal basis of 
 $\mathbb{C}^{2s+1}$, and the complex coefficients $a_{i}$ (some of which may 
 vanish) satisfy the normalization condition $\sum_{i} \vert a_{i}\vert^{2}=
 1$.
\end{fermionschmidt}
The {\em Slater number} of $\vert \psi(1,2)\rangle$ is then defined as 
 the number of non-zero coefficients $a_{i}$ appearing in the 
 decomposition of Eq.~(\ref{fermion0.5}).
Two possible cases can occur:\\

\noindent {\bf Slater Number $=1$}. In this situation the state
 $\vert \psi(1,2)\rangle$ has the form of a single Slater determinant:
\begin{equation}
\label{fermion1}
 \vert \psi(1,2)\rangle = \frac{1}{\sqrt{2}}\,[\vert 1\rangle_{1}\otimes
 \vert 2\rangle_{2} - \vert 2\rangle_{1}\otimes \vert 1\rangle_{2}\,]
\end{equation}

\noindent Since the state has been obtained by antisymmetrizing the product of
 two orthogonal states, $\vert 1\rangle$ and $\vert 2 \rangle$, it must
 be considered as non-entangled according to our previous
 criterion of Theorem~\ref{factorizabilityidentical2}.
The reduced single-particle statistical operators of each particle
 (it does not really matter
 which one we consider since, due to symmetry considerations, they
 are equal) and their von Neumann entropy (expressed in base $2$) are:
\begin{equation}
\label{fermion2}
\rho^{(1\:or\:2)} = \frac{1}{2}\,[\, \vert 1 \rangle \langle 1
 \vert + \vert 2 \rangle\langle 2 \vert \,]
\end{equation}
\begin{equation}
\label{fermion3}
 S(\rho^{(1\:or\:2)}) \equiv - \textrm{Tr} \,[\, \rho^{(1\:or\:2)} 
 \log \rho^{(1\:or \:2)}\,] = 1
\end{equation}

It should be obvious that we cannot pretend that the operator
 $\rho^{(1\:or\:2)}$ of Eq.~(\ref{fermion2}) describes the properties of 
 {\em precisely} the first or of the second particle of the system,
 due to their indistinguishability.
Accordingly, in this case, the value $S(\rho^{(1\:or\:2)})=1$
 correctly measures only the unavoidable uncertainty concerning the quantum 
 state to attribute to each of the two identical physical subsystems and
 it has nothing to do with any uncertainty arising from any actual
 form of entanglement.\\

\noindent {\bf Slater number $>1$}. In this case the state
 $\vert \psi(1,2)\rangle$ is written in term of more than one single Slater
 determinant and, consequently, it must be considered as a truly
 entangled state.
In fact, according to our criterion of Theorem~\ref{factorizabilityidentical2},
 there is no way to obtain $\vert \psi(1,2)\rangle$ by antisymmetrizing
 the tensor product of two orthogonal states.
Moreover the reduced single-particle statistical operators and their
 associated von Neumamn entropy are:
\begin{equation}
\label{fermion4}
\rho^{(1\:or\:2)} = \sum_{i=1}^{(2s+1)/2} \frac{\vert a_{i}\vert^{2}}{2}
 \,[  \vert 2i-1 \rangle \langle 2i-1 \vert + \vert 2i \rangle \langle
 2i\vert\,]
\end{equation}
\begin{equation}
\label{fermion5}
 S(\rho^{(1\:or\:2)}) = -\sum_{i=1}^{(2s+1)/2} \vert a_{i}\vert^{2}
 \log \frac{\vert a_{i}\vert^{2}}{2} =
 1-\sum_{i=1}^{(2s+1)/2} \vert a_{i}\vert^{2} \log \vert a_{i}\vert^{2} > 1
\end{equation}

In this case the fact that the von Neumann entropy is strictly greater than one
 correctly measures both the uncertainty deriving from the 
 indistinguishability of the particles and the one connected to the genuine 
 entanglement of the state. \\

The previous two cases are summarized in the following Theorem:
\begin{measurefermion}
 \label{measurefermion}
 A state vector $\vert \psi(1,2) \rangle$ describing two identical
 fermions is {\bf non-entangled} iff its Slater number is equal to 1 or,
 equivalently, iff the von Neumann entropy of the one-particle reduced
 density operator $S(\rho^{(1\: or \:2)})$ is equal to $1$.
\end{measurefermion}
 

\subsection{The boson case}

The case of bipartite systems composed ot two identical bosons is 
 slightly more articulated and subtle than the fermionic case.
Let us start, as before, by considering the bosonic Schmidt
 decomposition of an arbitrary state vector $\vert \psi(1,2) \rangle$
 belonging to the symmetric manifold ${\cal S}(\mathbb{C}^{2s+1}\otimes
 \mathbb{C}^{2s+1})$ and describing two identical bosons:
\begin{bosonschmidt}
 \label{bosonschmidt}
 Any state vector
 describing two identical $s$-spin boson particles $\vert \psi(1,2)
 \rangle$ and, consequently, belonging to the symmetric manifold
 ${\cal S}(\mathbb{C}^{2s+1}\otimes \mathbb{C}^{2s+1})$ can be written 
 as~\footnote{It is worth pointing out that the Schmidt decomposition of
 Eq.~(\ref{boson1}) is not always unique, as happens for the biorthonormal 
 decomposition of states describing distinguishable particles.}:
\begin{equation}
 \label{boson1}
 \vert \psi(1,2)\rangle = \sum_{i=1}^{2s+1} b_{i}\,
 \vert i \rangle_{1} \otimes \vert i \rangle_{2}\:,
\end{equation}
where the states $\left\{\, \vert i \rangle \right\}$, with
 $i=1,\dots, 2s+1$, constitute an orthonormal basis for $\mathbb{C}^{2s+1}$,
 and the real nonnegative coefficients $b_{i}$ satisfy the normalization 
 condition $\sum_{i}b_{i}^{2}= 1$.
\end{bosonschmidt}
The {\em Schmidt number} of the state of Eq.~(\ref{boson1}) is defined, as
 usual, as the number
 of non-zero coefficients $b_{i}$ appearing in the decomposition and
 the following cases must be analyzed:\\

\noindent {\bf Schmidt number $=1$}. In this case the state is factorized,
 i.e., 
 $\vert \psi(1,2) \rangle= \vert i^{\star} \rangle \otimes \vert i^{\star}
 \rangle$, and it describes two identical bosons in the same state 
 $\vert i^{\star} \rangle$.
It is obvious that such a state must be considered as non-entangled since 
 one knows precisely the properties of both constituents and, consequently,
 no uncertainty remains about which particle has which property.
This fact perfectly agrees with the von Neumann entropy of the single-particle
 reduced statistical operators $S(\rho^{(1 \,or\,2)})$ being null. \\

\noindent{\bf Schmidt number $=2$}. According to Eq.~(\ref{boson1}), the most
 general state with Schmidt number equal to $2$ has the following form:
\begin{equation}
 \label{boson4.3}
 \vert \psi(1,2) \rangle= b_{1}\vert 1 \rangle_{1} \otimes \vert 1\rangle_{2}
 + b_{2}\vert 2 \rangle_{1} \otimes \vert 2\rangle_{2},
\end{equation}
where $b_{1}^{2}+b_{2}^{2}=1$.

We can now distinguish two subcases, depending on the values of the
 positive coefficients $b_{1}$ and $b_{2}$. 
If they are equal, that is $b_{1}\!=\!b_{2}=\!1/\sqrt{2}$, the following
 Theorem holds:
\begin{bequalb}
 \label{bequalb}
The condition $b_{1}=b_{2}=1/\sqrt{2}$ is necessary and sufficient in order
 that the state $\vert \psi(1,2) \rangle= b_{1}\vert 1 \rangle_{1} \otimes
 \vert 1\rangle_{2} + b_{2}\vert 2 \rangle_{1} \otimes \vert 2\rangle_{2}$
 can be obtained by symmetrizing the factorized product of two orthogonal
 states.
\end{bequalb}
In this situation, and in full accordance with our
 Theorem~\ref{factorizabilityidentical3}, one must consider this state as
 non-entangled since it is possible to
 attribute well definite state vectors to both particles (of course, as usual, 
 we cannot say which particle is associated to which state due to their 
 indistinguishability). 
Moreover the von Neumann entropy of the reduced statistical operators
 $S(\rho^{(1\:or\:2)})$ is equal to $1$ measuring, as happened 
 in the fermion case with the state of Eq.~(\ref{fermion1}), 
 only the uncertainty arising from the indistinguishability of the particles. 

On the contrary, when the two coefficients are different, that is
  $b_{1}\neq b_{2}$, the following Theorem holds:
\begin{bdifferentb}
 \label{bdifferentb}
The condition $b_{1}\neq b_{2}$ is necessary and sufficient in order that
 the state $\vert \psi(1,2) \rangle= b_{1}\vert 1 \rangle_{1} \otimes
 \vert 1\rangle_{2} + b_{2}\vert 2 \rangle_{1} \otimes \vert 2\rangle_{2}$
 can be obtained by symmetrizing the factorized product of two
 non-orthogonal states.
\end{bdifferentb}
According to our original criterion, this state must be considered
 as a truly entangled state since it is impossible to attribute to both 
 particles definite properties.
Moreover the von Neumann entropy of the reduced statistical operator 
 $S(\rho^{(1\:or\:2)})= -b^{2}_{1}\log b_{1}^{2} - b_{2}^{2}\log b_{2}^{2}$
 lies within the open interval $(0,1)$.
It correctly measures the uncertainty coming both from the 
 indistinguishability of the particles and from the entanglement, and it
 is strictly less than one because, in a measurement process, there is a
 probability greater than $1/2$ to find both bosons in the 
 same physical state ($\vert 1\rangle$ or $\vert 2 \rangle$ depending
 whether $b_{1}>b_{2}$ or vice versa).\\

\noindent {\bf Schmidt number $\geq 3$}. In this situation the state is a
 genuine entangled one since
 it cannot be obtained by symmetrizing a factorized product of two
 orthogonal states~\footnote{In fact, if this would be true, the rank of the
 reduced density operator would be equal to two, in contradiction with the
 fact that a Schmidt number greater or equal to three implies a rank equal
 or greater to three.} and the von Neumann entropy
 of the reduced density operators is such that $S(\rho^{(1\,or\,2)})
 \in (0, \log(2s+1)]$. \\

In accordance with our analysis we can formulate the following Theorem 
 which supplies us with a unified and unambiguous criterion for
 detecting the entanglement in the boson case:
\begin{measureboson}
 \label{measureboson}
 A state vector $\vert \psi(1,2) \rangle$ describing two identical
 bosons is {\bf non-entangled} iff either its Schmidt number is 
 equal to $1$, or the Schmidt number is equal to $2$ {\em and}  the 
 von Neumann entropy of
 the one-particle reduced density operator $S(\rho^{(1\: or \:2)})$ is
 equal to $1$. 
Alternatively, one might say that the state is {\bf 
 non-entangled} iff either its von Neumann entropy is equal to $0$, or it is 
 equal to $1$ and the Schmidt number is equal to $2$.
\end{measureboson}
It is clear from the previous analysis, differently from what 
 happens in the case of two identical fermions, that the consideration
 of the Schmidt number alone (or of the von Neumann entropy) to detect 
 entanglement for a pair of bosons is inappropriate.
In fact there exist states with Schmidt number equal to $2$, or with
 von Neumann entropy equal to $1$, which can be non-entangled as well as 
 entangled.
Therefore, the only consistent way to overcome this problem turns out to 
 be that of considering the two criteria together, as clearly stated
 in the Theorem~\ref{measureboson}.


\section{Conclusions}

In this paper we have reviewed in general the problem of
 deciding whether a state describing a system of two identical
 particles is entangled or not. 
Two equivalent criteria have been exhibited: the first~\cite{gmw,gm}, in the
 spirit of the founder fathers of Quantum Mechanics, is based on the 
 possibility of attributing a complete set of objective properties to both 
 constituents (that is, a definite state vector) while the second~\cite{gm2} 
 is based on the consideration of {\em both} the Slater-Schmidt number of 
 the fermionic and bosonic analog of the Schmidt decompositions of the states 
 describing the system {\em and} of the von Neumann entropy of the reduced 
 statistical operators.


\end{document}